\documentclass[preprint, superscriptaddress, aps]{revtex4}
\usepackage[english]{babel}
\usepackage[utf8]{inputenc}
\usepackage{hyperref}
\usepackage{amsmath}
\usepackage{amssymb}
\usepackage{amsfonts}
\usepackage{graphicx}

\begin{document}
	
		\begin{center}
		\section*{\Large Strong suppression of shot noise in a feedback-controlled single-electron transistor. \vspace{1cm}}
		
		\large
		T.~Wagner$^1$, P. Strasberg$^2$, J. C. Bayer$^1$, E. P. Rugeramigabo$^1$, \\ T. Brandes$^2$ and R.J. Haug$^1$
		\vspace{1cm}
		
		\textit{{\small $^1$Institut für Festkörperphysik, Leibniz Universität Hannover, D-30167 Hannover, Germany}\vspace{-0.3cm}}
		\textit{{\small $^2$Institut für Theoretische Physik, Hardenbergstr.~36, TU Berlin, D-10623 Berlin, Germany \vspace{1.5cm}}}
	\end{center}

	\begin{center}
		\section*{\large Abstract}
		
		\begin{changemargin}{1.5cm}{1.5cm}{\fontsize{10}{1}\selectfont Feedback control of quantum mechanical systems is rapidly attracting attention not only due to fundamental questions about quantum measurements \cite{QMWiseman} but also because of its novel applications in many fields in physics. Quantum control has been studied intensively in quantum optics \cite{QMWiseman, QOSerafini} but recently progress has been made in the control of solid-state qubits \cite{NVBlok, SRVijay, ECYacoby} as well. In quantum transport only a few active \cite{TNChida, FEHofmann, SEKoski} and passive \cite{MFFricke, TTThierschmann, OCKoski} feedback experiments have been realized on the level of single-electrons, though theoretical proposals \cite{FCBrandes,CPKiesslich, CFEmarya} exist.}
			
		{\fontsize{10}{1}\selectfont
		Here we demonstrate the suppression of shot noise in a single-electron transistor, using an exclusively electronic closed-loop feedback to monitor and adjust the counting statistics \cite{FCNazarov, RTDLu, CMBylander, CSGustavsson, UOFlindt, SRFricke, TNChida}. With increasing feedback response we observe a stronger suppression and faster freezing of charge current fluctuations. Our technique is analog to the generation of squeezed light with in-loop photodetection \cite{QMWiseman,  SLMachida, ODWalker} as used in quantum optics. Sub-Poisson single-electron sources will pave the way for high precision measurements in quantum transport similar to its optical or opto-mechanical \cite{QSWollman} equivalent.}
		\end{changemargin} 	
	\end{center}
	\newpage
	
	\normalsize 
	A current of electrons passing through a potential barrier follows a Poisson process, characterized by white noise. This so-called shot noise originates from the quantization of the electric charge $e$ and has a current spectral density $S_{P} = 2e\langle I \rangle$ proportional to the average current $\langle I \rangle$. Its occurrence was first postulated in vacuum diodes by W.~Schottky in 1918 \cite{SNSchottky} and is the dominant noise source in present day mesoscopic conductors \cite{SNBlanter}.
	
	Correlations between the electrons cause deviations in the spectral density $S$ from the Poissonian limit $S_{P}$. A single-electron transistor (SET) only allows sequential transfer of electrons due to Coulomb blockade, which leads to a minimum Fano factor $F_{SET} = S/S_P = 0.5$ for symmetric tunnel-couplings\cite{CSGustavsson, SNBlanter}. Shot noise measurements therefore give a deep insight into the internal physical processes of a system \cite{NSLandauer}. 
	
	For a further suppression of shot-noise in single-electron devices, additional time-correlations between the tunneling events have to be imposed. One approach is to drive the tunneling process periodically, capturing and releasing one electron per cycle as done by single-electron pumps and turnstiles \cite{SEPekola}. But these open-loop systems are not robust against stochastic fluctuations of the underlying quantum mechanical tunneling process, leading to an inevitable increase of shot noise \cite{SRFricke}.
	
	Our approach here is the implementation of a measurement-based, active closed-loop feedback control \cite{QMWiseman} to stabilize the random charge fluctuations in an SET. The feedback loop monitors the single-electron tunneling in real-time with a charge detector and feeds back periodically the deviation from a target rate to a control gate, to speed up or slow down the process. Accordingly, more or less electrons will be transferred in the next interval to compensate the deviations.
		
\begin{figure}
	\centering
	\includegraphics[width=0.6 \textwidth, bb=0 -20 260 390]{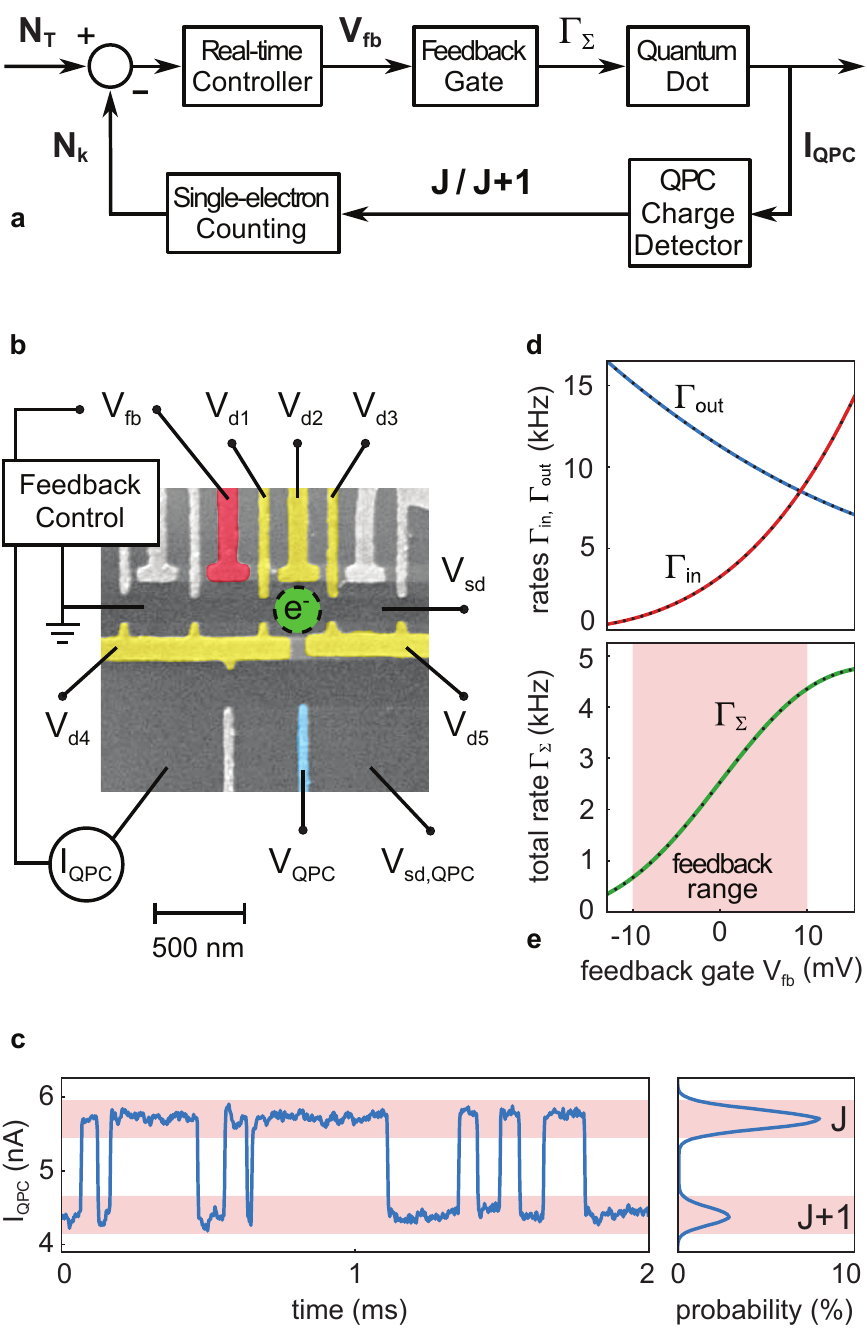} 
	\caption{{\textbf{Feedback set-up.} \textbf{a,} Block diagram of implemented negative feedback loop showing the involved process components and their dependencies with each other. \textbf{b,} SEM image of the sample structure. The quantum dot gates V$_{d1}$ to V$_{d5}$ are highlighted golden, the quantum point contact (QPC) charge detector V$_{qpc}$ bluish and the feedback gate V$_{fb}$ reddish. The gray gates are not used in this experiment and unbiased. The visible gap between dot and detector is closed electrostatically and enhances the coupling between both. \textbf{c,} Typical time-resolved charge detection measurement with adjacent QPC. The two different charging states J and J+1 of the quantum dot are well separated. \textbf{d,} Feedback gate V$_{fb}$ dependent rates for tunneling in $\Gamma_{in}$ and out $\Gamma_{out}$ of the quantum dot. \textbf{e,} The resulting total tunneling rate $\Gamma_{\sum}=\Gamma_{in}\Gamma_{out}/(\Gamma_{in}+\Gamma_{out})$. In the red highlighted quasi linear range the feedback experiments were carried out.}}
	\label{fig:sample}
\end{figure}

	Fig.~\ref{fig:sample} shows the corresponding block diagram, our sample structure and the controlled physical quantities. The SET consists of a quantum dot, and to ensure a directed tunneling \cite{CSGustavsson, UOFlindt} we apply a bias voltage V$_{sd,dot} = 1.5$~mV between source and drain. The single-electron charge detector is formed by a coupled quantum point contact (QPC) \cite{CSGustavsson, UOFlindt}. A snapshot of a time-resolved detector signal can be seen in Fig.~\ref{fig:sample}c. Whenever the number of electrons on the quantum dot changes, there is a discrete jump visible. From the associated waiting time distributions, the tunneling rates for source $\Gamma_{in}$ and drain $\Gamma_{out}$ can be extracted \cite{CSGustavsson, UOFlindt}. By varying the feedback gate voltage V$_{fb}$ we are able to alter the rates as shown in Fig. \ref{fig:sample}d. Two different effects determine the dependence. On the one hand, a more negative gate voltage shifts the transport level from a resonance with drain to a resonance with source, causing the opposite slopes. On the other hand the tunnel barriers are affected differently by the applied gate voltage, which results in different gradients. The total tunneling rate $\Gamma_{\sum}$ has a quasi-linear range within our feedback experiments (Fig. \ref{fig:sample}e).

\begin{figure}
	\centering
	\includegraphics[width=0.9 \textwidth]{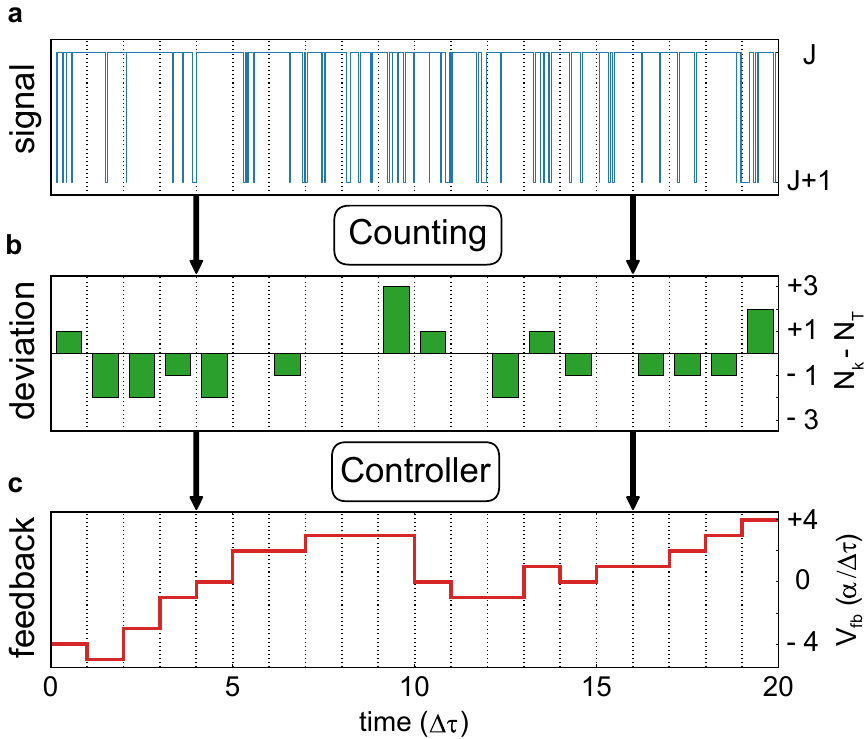} 
	\caption{\textbf{Detailed feedback processing.} \textbf{a,} First the QPC charge signal is digitized and serves as feedback input. \textbf{b,} Afterwards the number of tunneled electrons $N_{k}$ are counted within the feedback window $\Delta \tau$ (here  1~ms) and compared to the specified target number $N_{T}$ (here  3). \textbf{c,} Finally, the feedback output is achieved by adjusting the feedback gate voltage $V_{fb}$ relative to the deviation $N_{k} - N_{T}$ (eq. \ref{eq:feedback}).}
	
	\label{fig:feedback}
\end{figure}

	In Fig. \ref{fig:feedback} the feedback signal processing is shown in detail. The digitized time-resolved counting signal serves as input (Fig. \ref{fig:feedback}a). We compare the number of transferred electrons $N_{k}$ within a constant time window $ \Delta \tau $ with a defined target number $ N_{T} $ (Fig. \ref{fig:feedback}b). Finally, the output correction is achieved by changing the feedback gate voltage $ V_{fb} $ linearly to the deviation (Fig. \ref{fig:feedback}c)
	\begin{equation}
		V_{fb, k+1} = V_{fb, k} - \frac{\alpha}{\Delta \tau} ~ (N_{k} - N_{T})
		\label{eq:feedback}.
	\end{equation}
	To vary the feedback response, we can either change the feedback factor $\alpha$ or the window $\Delta \tau$. The target rate is defined as $ \Gamma_{T} = N_{T} / \Delta \tau $. For the feedback mechanism the gate dependence of the total rate $\Gamma_{\sum}$ does not matter, because we only vary the feedback voltage V$_{fb}$ relatively. However, the rate characterization is important for setting reasonable feedback parameters and limits. The target rate $\Gamma_{T}$ is always selected to be centered in the quasi-linear range. Outside of it the feedback loop becomes unstable. In particular, with increasing feedback response $\alpha / \Delta \tau$ the feedback gate variations become larger and the voltage might run out of the quasi-linear range (Fig. \ref{fig:sample}e). To avoid this, we limit the feedback to the quasi-linear range and if the output $V_{fb, k+1}$ lies outside, we set the boundary value instead.

\begin{figure}
	\centering
	\includegraphics[width=0.8 \textwidth]{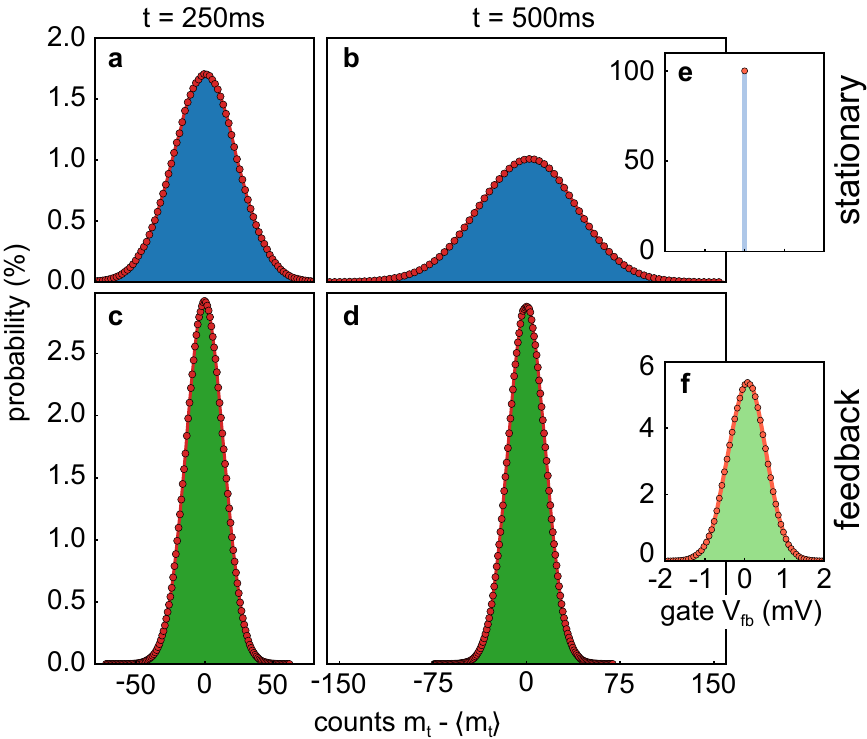} 
	\caption{\textbf{Stationary and feedback distributions.} \textbf{a, b,} The counting distributions of the stationary quantum dot (blue) are shown for two different integration times $t$. A strong temporal broadening of the charge fluctuations is visible. \textbf{c, d,} In comparison the distributions of the feedbacked quantum dot (green) are clearly suppressed and temporally frozen. \textbf{e, f,} At the same time the distribution of applied gate voltages $V_{fb}$ transforms from a single-valued Delta-peak into a broad distribution. The target rate was in both cases $\Gamma_{T} = 2$~kHz and the feedback parameters were $\Delta \tau=5$~ms, $\alpha = 1 \cdot 10^{-5}$~Vs.}
	\label{fig:histograms}
\end{figure}

	We evaluate the influence of the feedback by extracting the full counting statistics \cite{FCNazarov, CSGustavsson, UOFlindt}. Therefore the charge signal is recorded and divided into equal time slots. The length of the slots should not be confused with the feedback window $\Delta \tau$  and for a clear distinction we call it integration time $t$. While $\Delta \tau$  is of experimental relevance, the typical much larger integration times are only used in the statistical analysis. By counting the number of tunneled electrons $m_{t}$ in each slot, we finally obtain the counting distribution, describing the charge fluctuations on the time scale of $t$. Fig. \ref{fig:histograms} compares the charge fluctuations of the SET for the stationary case (blue) without feedback and for the dynamic case (green) with activated feedback loop. From the different counting distributions, we can already qualitatively recognize the two characteristic effects of the feedback. The first one is the suppression of the shot noise (similarly to the sub-Poissonian photon number statistics of squeezed light) and the second one is the temporal freezing of the fluctuations. Without feedback, a fast broadening happens with increasing integration time because the tunneling events occur randomly and larger deviations from the mean $\langle m_{t} \rangle$ become more likely for longer integration times. The feedback loop prevents this behavior and the charge distributions become integration time independent. So deviations from the target number on short time scales are compensated for sufficiently long integration times, resulting in a frozen counting distribution. At the same time the variations of the feedback gate increase and the distribution of applied gate voltages $V_{fb}$ transforms from a single-valued Delta-peak into a broad distribution.
	
\begin{figure*}
	\includegraphics[width=1.0 \textwidth]{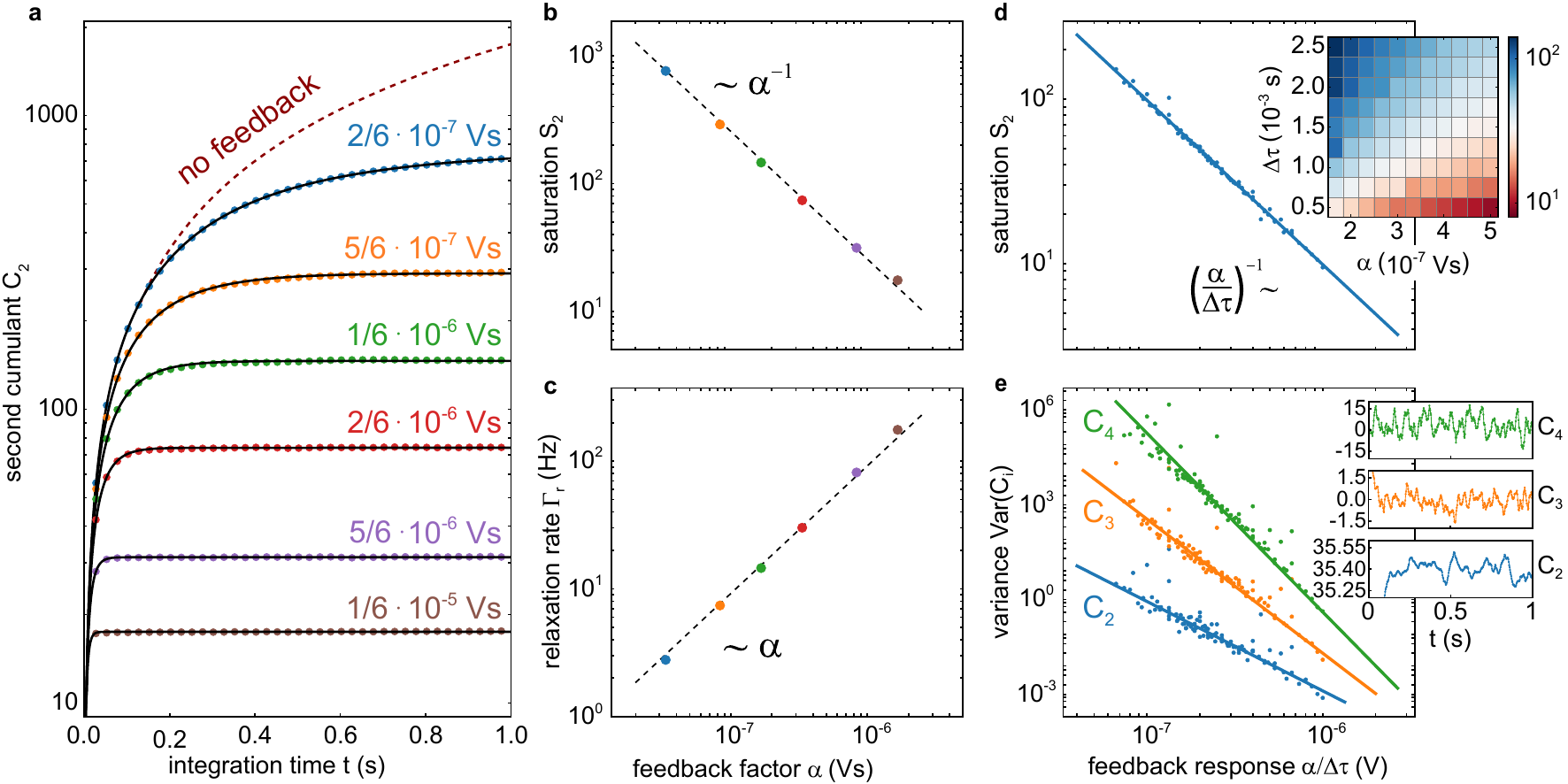} 
	\caption{\textbf{Feedback dependent cumulants of the charge fluctuations} \textbf{a,} The second cumulant $C_{2}$ is plotted as a function of integration time $t$ for different feedback factors $\alpha$. All curves show the feedback characteristic freezing for sufficient long integration times and with stronger feedback factor $\alpha$ the suppression of shot noise increases. The red dashed line indicates the shot noise $C_{2}^{st} = 0.5 \Gamma_{T}t$ of a stationary quantum dot with symmetric tunneling rates. The black dashed lines are determined by fitting the data with eq. 2. \textbf{b, c,} Feedback factor $\alpha$ dependence of the extracted fit parameters. The dashed lines show an inverse proportionality of the saturation value $S_2 \sim \alpha^{-1}$ and a direct proportionality for the relaxation rate $\Gamma_r \sim \alpha$. \textbf{d,} The colorplot shows the saturation value $S_2$ for different feedback factors $\alpha$ and windows $\Delta \tau$. A transition from a weak feedback region to a strong region is clearly visible. Plotted as a function of the feedback response $\frac{\alpha}{\Delta \tau}$ the saturation value $S_{2}$ reveals a linear dependence. \textbf{e,} All cumulants show fluctuations around their saturation value. The variance of those fluctuations is strongly suppressed with increasing feedback response and follows the empirical power law $Var(C_m) \sim (\frac{\alpha}{\Delta \tau})^{-m \cdot \sqrt{2}}$.}
	\label{fig:cumulant_dependency}
\end{figure*}

	A quantitative description of the feedback efficiency is obtained from the cumulants $C_j^{fb}$ of the counting distributions as a function of integration time $t$. The second cumulant $C_{2}^{fb}$ (variance) is plotted in Fig.~\ref{fig:cumulant_dependency}a for a wide range of feedback factors $\alpha$. Each curve results from the evaluation of a 60~minute trace, including 1,440,000 feedback corrections and about 12,600,000 tunnel events. This ensures an excellent statistical accuracy even for long integration times. The feedback window was kept at $\Delta \tau=2.5$~ms and the target rate has been set to $\Gamma_{T}=3.5$~kHz. All curves show a clear saturation for sufficiently long integration times, indicating the characteristic freezing. With increasing feedback factor $\alpha$ the saturation value becomes smaller, which corresponds to a stronger suppression of shot noise. For a direct comparison we also plotted $C_{2}^{st} = 0.5 \Gamma_T t$ (red dashed line), corresponding to the minimum observable shot noise in a stationary SET.
	
	The dependence of the second cumulant on integration time corresponds to the equation
	\begin{equation}
		C_{2}^{fb}(t) = S_{2} \ (1 - e^{-\Gamma_{r} \cdot t})
		\label{eq:2cumulant}
	\end{equation}	
	which was first derived for a single barrier with continuous feedback \cite{FCBrandes} and which we extended to our applied discrete feedback scheme (see Supplementary). $S_{2} $ is the saturation value for long integration times $t$ and the relaxation rate $ \Gamma_{r} $ describes how fast the fluctuations freeze. The two parameters are effective values that we gain by fitting eq. \ref{eq:2cumulant} to the measured curves (black lines in Fig. \ref{fig:cumulant_dependency}a). From the fits we find that the saturation value is inversely proportional $S_{2} \sim \alpha^{-1}$ (Fig. \ref{fig:cumulant_dependency}b) and the relaxation rate is directly proportional $\Gamma_{r} \sim \alpha$ (Fig. \ref{fig:cumulant_dependency}c) to the feedback factor. So not only the shot noise suppression is stronger but also the freezing occurs more rapidly with increasing feedback factor. The measured dependencies of the second cumulant agree with our theoretical calculations in the Supplementary. Although the derivation of eq 2 is based on a single barrier, we find for our measured dot system exactly the same behavior. This indicates the robustness of the discrete feedback scheme.
	
	For the strongest applied feedback $\alpha=1/6 \cdot 10^{-5}$~Vs with $\Gamma_{T}=3.5$~kHz we extracted $S_{2} = 17.5$. This gives for the Fano factor $F_{fb} = \frac{C_{2}}{\Gamma_T t} = 0.005$ at $t=1$~s and corresponds to a shot noise suppression of 23~dB. The feedback loop bandwidth is characterized by  $\Gamma_r$, which is in the present case 176~Hz.
	
	In a second experiment, we studied the influence of the feedback window $\Delta \tau$ by varying it in addition to the feedback factor $\alpha$. For every combination we recorded a 5~minute trace with a target rate of $4$~kHz, which corresponds to 1,200,000 tunnel events. The dependence of the saturation value $S_2$ is shown as colorplot inset in Fig. \ref{fig:cumulant_dependency}d. Clearly, the transition from a weakly feedbacked system on the top left to a strongly feedbacked one on the bottom right is visible, with S${_2}$ determined by the response response $\alpha / \Delta \tau$. By varying the feedback window at a constant target rate, we actually change the target number $N_{T} =  \Gamma_T \Delta \tau$ and can assume $S_{2} \sim N_{T}/\alpha $ (as theoretically proven in the Supplementary). With the extracted relations we can express the feedback dependent Fano factor $F_{fb} = \frac{C_{2}}{\Gamma_T t} \sim \frac{\Delta \tau}{\alpha} \frac{1}{t}$ for $t > \Gamma_r^{-1}$. It is independent of the target rate and decreases with integration time as a result of the frozen charge fluctuations.
	
	Higher cumulants show fluctuations around zero as a function of integration time (inset of Fig. \ref{fig:cumulant_dependency}e) instead of freezing at feedback dependent values. Note, however, that without feedback all cumulants would grow linearly in time; a behaviour which is completely prevented by the feedback. On a closer inspection also the second cumulant fluctuates weakly around its saturation value. We attribute this fluctuating behavior mainly to the statistical error due to the finite experimental counting samples. Nevertheless a strong suppression of the amplitudes is observable with increasing feedback response. To characterize the feedback influence we calculated the variance of the cumulants in the range $t = [0.25~s,~1.0~s]$ and plotted the results in Fig. \ref{fig:cumulant_dependency}e against the feedback response $\alpha / \Delta \tau$. It turns out that a power law exists for all extracted quantities with increasing exponent for higher orders. The solid lines indicate a good agreement with the empirical power law $Var(C_m) \sim (\frac{\alpha}{\Delta \tau})^{-m \cdot \sqrt{2}}$. So with increasing feedback response all cumulants become smaller and less noisy as a function of time. Therefore the frozen counting distributions can be well approximated by Gaussian functions.
	
	For very strong feedback responses $\alpha / \Delta \tau$ the feedback gate voltage exceeds the boundaries of the quasi linear range, and the second cumulant exhibits a weak increase with integration time. Nevertheless, the suppression is still large, which shows that the feedback loop is very robust even under stronger experimental restrictions. Extended gate dependencies, which vary the tunneling barriers and keep the transport level constant, would allow stronger feedback responses.
	
	For practical applications larger currents would be desirable. This could be achieved by operating many SETs in parallel. With individual charge detectors, we can count the transferred electrons $N_{k,i}$ separately but apply a feedback being proportional to the total number  $N_{k} = \sum{N_{k,i}}$. It is not necessary to control every dot individually, the fluctuations can be stabilized by feedbacking just one or a few dots to reduce the device complexity. Furthermore, the operation is not affected by different tunneling rates of the individual dots, making the up-scaling tolerant against unavoidable fabrication differences. Complementary to the parallelization higher currents result from higher individual tunneling rates. These can be resolved by radio-frequency charge detectors, with bandwidths above 10~MHz already shown \cite{RTDLu, CMBylander, RFQPCCassidy}.
	
	For the metrological redefinition of the unit ampere a highly accurate current source is necessary, directly connecting an external frequency $f$ with the output current $I{=}ef$. The initially mentioned single-electron pumps are promising candidates \cite{SEPekola}. But the unavoidable fluctuations of the generated output current lead to an increase of shot noise with integration time. Also the theoretical discussed suppression of fluctuations in serial pumps \cite{MFFricke}, due to a passive mesoscopic feedback, only reduces the internal fluctuations between the pumps. For an accurate current definition, always the output fluctuations must be detected and accounted for \cite{SRFricke, SRWulf}. The feedback controlled SET current $I_{SET} = e N_T \Delta \tau^{-1}$ also fulfills the above relation, for the redefinition of the ampere. But unlike the pumps, it stabilizes the fluctuations of the output current, which therefore becomes more precise with increasing integration time. We would like to motivate that the feedback technique is generic for single-electron sources and not limited to SETs. As already suggested in L. Fricke et al. \cite{MFFricke}, it could be implemented to correct erroneous pumping events in serial pumps \cite{SRFricke, SRWulf}. 
	
	To conclude, we successfully implemented an exclusively electronic feedback loop to stabilize the single-electron tunneling process through a SET and analyzed the influence of different feedback parameters with the full counting statistics. We observed a strong suppression of shot noise with increasing feedback response $\frac{\alpha}{\Delta \tau}$ and the freezing of the charge fluctuations on sufficiently long time scales. Our results suggest single-electron sources with integrated feedback logic as high-precision on-chip current sources. Furthermore, feedback controlled single-electron source are an ideal platform to test non-equilibrium thermodynamics at the nanoscale (Maxwell's demon)\cite{SEKoski, TMStrasberg, OCKoski} or to build efficient thermoelectric devices (energy harvesters)\cite{TESanchez, TTThierschmann}.
	
	{}
	\vspace*{1cm}
	\section{Methods}
	All measurements were carried out on a low-noise DC transport setup in a $^{4}$He cryostat at 1.5~K. Our sample is based on a GaAs/AlGaAs heterostructure, forming 100~nm below the surface a two-dimensional electron gas (2DEG). The 2DEG charge carrier density is $n_{e}=2.4~{\cdot}~10^{11}$~cm$^{-2}$ and the mobility is $\mu_{e}=5.1~{\cdot}~10^5$~cm$^{2}$/Vs. We define the quantum dot (QD) and quantum point contact (QPC) with metallic topgates (7~nm Cr, 30~nm Au), processed with optical and electron beam lithography. The structure is formed by applying negative voltages to the gates, depleting the electron gas below it. The feedback gate was filtered by 1~MHz low-pass filter at room  temperature. All other gates, as well as QD and QPC source were filtered by a 10~Hz low-pass filter. The detector current was amplified outside the cryostat with a low-noise FEMTO transimpedance amplifier (100 MV/A gain, 100kHz bandwidth), connected to the QPC drain by a 25~pF low~capacity coax line [1]. Subsequent signal processing and feedback control was fully done by a programmable Adwin Pro 2 real-time controller, consisting of a 333~MHz ADSP (T11), 18~bit ADC card (AIn F-8/18) and 16~bit DAC card (AOut 8/16). The processing rate was $\Gamma_{s}$=400kHz. To increase the output resolution a voltage divider 1:30 was used. Time delays were negligible because the inverse RC time $\tau_{RC}^{-1}$ and sampling rate are large compared to the target rate $\Gamma_{T} \ll \tau_{RC}^{-1}, \Gamma_{s} $. In addition the raw QPC signal was recorded for the processing validation and later statistical analysis.
	\\
	\\
	$[1]$ {\small Maire, N., Hohls, F., Lüdtke T., Pierz. K \& Haug, R. J. Noise at a Fermi-edge singularity in self-assembled InAs quantum dots. \textit{Phys. Rev. B} \textbf{75}, 233304 (2007).}
		
	\newpage
	
	\section{Acknowledgments}
	We thank Géraldine Haack for the valuable discussions. This work was financially supported by the DFG GRK 1991, QUEST (T.W, J.C.B., E.R., R.J.H.) and DFG SFB 910, GRK 1558 (P.S., T.B.).
	
	\section{Author contributions}
	T.W. carried out the experiments, analyzed the data and wrote the manuscript. J.C.B. and T.W. fabricated the device. E.R. provided the wafer material. P.S and T.B. provided theory support and supplied Supplementary material. R.J.H. supervised the research. All authors discussed the results and contributed to editing the manuscript.
	
	\section{Additional Information}
	Supplementary information is available in the online version of the paper. Reprints and permission information is available online at www.nature.com/reprints. Correspondence and requests for materials should be addressed to T.W. or R.J.H.
	
	\section{Competing financial interests}
	The authors declare no competing financial interests.
	
\newpage
~
\newpage
~
\newpage
~
\newpage

\begin{center}
	\section*{\Large Supplementary information for ``Squeezing of shot noise using feedback controlled single-electron tunneling''\vspace{1cm}}
	
	\large
	T.~Wagner$^1$, P. Strasberg$^2$, J. C. Bayer$^1$, E. P. Rugeramigabo$^1$, \\ T. Brandes$^2$ and R.J. Haug$^1$
	\vspace{1cm}
	
	\textit{{\small $^1$Institut für Festkörperphysik, Leibniz Universität Hannover, D-30167 Hannover, Germany}\vspace{-0.3cm}}
	\textit{{\small $^2$Institut für Theoretische Physik, Hardenbergstr.~36, TU Berlin, D-10623 Berlin, Germany \vspace{1.5cm}}}

\end{center}

We present a simplified theoretical treatment of the experiment: instead of considering the full quantum dot, we neglect any internal structure of it and treat the system as a single barrier. Our main result is a recursive relation for the cumulant generating function from which it is possible to obtain the steady state values of all cumulants. Furthermore, we solve for the full dynamical behaviour of the first and second cumulant. The theory developed here underlines the experimental results in the main text analytically.

The central object of our interest is the probability distribution $p_{k}(n_k)$ of having observed a \emph{total number} $n_k$ of tunneled electrons after a time $t_k = k\Delta\tau$, i.e., after $k$ feedback intervals. We begin by noting that Eq.~(1) in the main text can be recursively iterated such that 
\begin{equation}
\begin{split}
V_{fb,k+1}	&=	V_{fb,k} - \frac{\alpha}{\Delta\tau}(N_k-N_T) = V_{fb,k-1} -  \frac{\alpha}{\Delta\tau}(N_{k-1}-N_T) - \frac{\alpha}{\Delta\tau}(N_k-N_T) = \dots 	\\
&=	V_{fb,1} - \frac{\alpha}{\Delta\tau}(n_k-kN_T).
\end{split}
\end{equation}
Here, $V_{fb,1}$ denotes the initial voltage and $n_k$ is the total number of tunneled electrons after time $t_k$. The experiment is performed in a regime where the effective tunneling rate $\Gamma_\Sigma$ depends linear on the feedback gate voltage (see Fig.~1e), hence we can set 
\begin{equation}\label{eq feedback rates}
\Gamma_{\Sigma,k+1} = \Gamma_{\Sigma,0} + c V_{fb,k+1} = \Gamma_{\Sigma,1} - \frac{\alpha'}{\Delta\tau}(n_k-kN_T)
\end{equation}
where the constant $c$ corresponds to the slope of the curve shown in Fig.~1e and $\alpha' \equiv c\alpha$. Note that in order to have an always positive rate $\Gamma_{\Sigma,k+1}$ the feedback factor $\alpha'$ should be relatively small. Furthermore, the limit of a continuous feedback loop as in Ref.~\cite{BrandesPRL2010} can be reproduced by choosing an infinitesimal small feedback window, $\Delta\tau\rightarrow0$, while keeping $\alpha'/\Delta\tau$ and $kN_T$ finite. For finite $\Delta\tau$, however, it is \emph{a priori} not clear whether it is possible to reproduce the predictions of 
Ref.~\cite{BrandesPRL2010} even qualitatively. We will now show that this is indeed the case and thereby demonstrate the robustness of this kind of feedback loop. 

Because we are considering the case of a single barrier, we know that the number of tunneled particles within the $(k+1)$'th feedback interval $[k\Delta\tau,(k+1)\Delta\tau]$ obeys a Poisson process according to the rate $\Gamma_{\Sigma,k+1}$, 
\begin{equation}\label{eq Poisson process}
P(N_{k+1}) = \frac{(\Gamma_{\Sigma,k+1}\Delta\tau)^{N_{k+1}}}{N_{k+1}!} e^{-\Gamma_{\Sigma,k+1}\Delta\tau},
\end{equation}
where $N_{k+1} = n_{k+1}-n_k$ denotes the number of tunneled particles within that interval. Here, it should be noted that $\Gamma_{\Sigma,k+1}$ depends on $n_k$ through Eq.~(\ref{eq feedback rates}). With the help of Eq.~(\ref{eq Poisson process}) we can then link the probability $p_{k+1}(n_{k+1})$ to observe $n_{k+1}$ tunneled particles after time $t_{k+1}$ to $p_k(n_k)$ via the recursion relation 
\begin{equation}\label{eq recursion relation probabilities}
p_{k+1}(n_{k+1}) = \sum_{n_k=0}^{n_{k+1}} P(n_{k+1}-n_k) p_k(n_k). 
\end{equation}
Solving this with a given initial condition (we will assume below that $p_0(n_0) = \delta_{0,n_0}$ where $\delta_{n,m}$ denotes the Kronecker symbol) will in principle give us complete information about the stochastic process. However, in turns out to be more convenient to introduce the cumulant generating function $C_k(\chi) \equiv \ln \sum_{n_k=0}^\infty e^{in_k\chi}p_k(n_k)$. From~(\ref{eq recursion relation probabilities}) we then obtain another recursion relation 
\begin{equation}\label{eq recursion relation CGF}
C_{k+1}(\chi) = -\Delta\tau\left(\Gamma_{\Sigma,1} + k \frac{\alpha'}{\Delta\tau} N_T\right) (1-e^{i\chi}) + C_k[\chi-i\alpha'(1-e^{i\chi})].
\end{equation}
This can be proven by noting that 
\begin{equation}
\begin{split}
e^{C_{k+1}(\chi)}	&=	\sum_{n_k=0}^\infty e^{in_{k+1}\chi} \sum_{n_k=0}^{n_{k+1}} P(n_{k+1}-n_k) p_k(n_k)	\\
&=	\sum_{N_k=0}^\infty \sum_{n_k=0}^\infty e^{i(N_k+n_k)\chi} \frac{(\Gamma_{\Sigma,k+1}\Delta\tau)^{N_k}}{N_k!} e^{-\Gamma_{\Sigma,k+1} \Delta\tau} p_k(n_k)	\\
&=	\sum_{n_k=0}^\infty \exp\left\{-\Delta\tau\left[\Gamma_{\Sigma,1} - \frac{\alpha'}{\Delta\tau}(n_k - k N_T)\right](1-e^{i\chi})\right\} e^{in_k\chi} p_k(n_k)	\\
&=	\exp\left[-\Delta\tau\left(\Gamma_{\Sigma,1} + i\frac{\alpha'}{\tau}kN_T\right)(1-e^{i\chi})\right] \sum_{n_k=0}^\infty e^{in_k[\chi - i\alpha'(1-e^{i\chi})]} p_k(n_k)
\end{split}
\end{equation}
from which the desired result follows by taking the logarithm. By virtue of relation~(\ref{eq recursion relation CGF}) we can now start to compute the $\ell$'th cumulant $C_\ell^k \equiv \partial_{i\chi}^\ell C_k(\chi = 0)$ recursively. For the first cumulant, which equals the first moment, we obtain 
\begin{equation}
C_1^{k+1} = (1-\alpha') C_1^k + \Gamma_{\Sigma,1} \Delta\tau + \alpha' k N_T.
\end{equation}
This can be solved exactly and yields 
\begin{equation}
C_1^k = \frac{\Gamma_{\Sigma,1}\Delta\tau - \Gamma_{\Sigma,1}\Delta\tau (1-\alpha')^k + N_T \left[(1-\alpha')^k + \alpha' k-1\right]}{\alpha'}.
\end{equation}
Consequently, the steady state current is given by $\lim_{k\rightarrow\infty} \frac{C_1^k}{k} = N_T$, which was to be expected~\cite{BrandesPRL2010}. For the second cumulant (the variance) we obtain the recursion relation 
\begin{equation}
C_2^{k+1} = (1-\alpha')^2 C_2^k - \alpha' C_1^k + \Gamma_{\Sigma,1}\Delta\tau + \alpha' k N_T. 
\end{equation}
Assuming for simplicity that the first moment has reached its steady state, we obtain the simpler relation 
\begin{equation}
C_2^{k+1} = (1-\alpha')^2 C_2^k + \Gamma_{\Sigma,1}\Delta\tau. 
\end{equation}
This can be solved easily again and we obtain 
\begin{equation}
C_2^k = N_T\frac{1-[(1-\alpha')^2]^k}{1-(1-\alpha')^2} \underset{\alpha'\ll 1}{\approx} \frac{N_T}{2\alpha'}\left[1-(1-2\alpha')^k\right].
\end{equation}
Thus, the convergence of the variance to its steady state $\lim_{k\rightarrow\infty} C_2^k \approx \frac{N_T}{2\alpha'}$ is exponential in time and agrees with the prediction of Ref.~\cite{BrandesPRL2010} in the limit $\Delta\tau\rightarrow0$. In fact, in the experiment performed the feedback window $\Delta\tau$ is relatively small compared to the total time window over which the evolution is observed. It thus makes sense to consider the case of $N\gg1$ feedback intervals and $\Delta\tau \sim N^{-1} \ll 1$. Then, we have 
\begin{equation}
(1-2\alpha')^k \approx \exp\left(-\frac{2\alpha'}{\Delta\tau} t\right)
\end{equation}
where $t = k\Delta\tau$. After using the definition of the target rate, $\Gamma_T = \Delta\tau N_T$, we thus obtain 
\begin{equation}
C_2^k \approx C_2(t) = \frac{N_T}{2\alpha'} \left[1 - \exp\left(-2\frac{\alpha'}{N_T}\Gamma_T t\right)\right],
\end{equation}
which has to be compared with the experimental results, see Fig.~4, and Eq.~(2) with $S_2 \equiv \frac{N_T}{2\alpha'}$ and $\Gamma_r \equiv 2\frac{\alpha'}{N_T}\Gamma_T$. In principle, it is possible to go on with this procedure. For instance, the steady state value of the third cumulant (proportional to the skewness of the distribution) is given by 
\begin{equation}
\lim_{k\rightarrow\infty} C_3^k = N_T \frac{1-2\alpha'}{\alpha'(-6 + 9\alpha' - 5\alpha'^2 + \alpha'^3)} \underset{\alpha'\ll 1}{\approx} -\frac{N_T}{6\alpha'},
\end{equation}
which then again agrees with Ref.~\cite{BrandesPRL2010} for the case of continuous feedback control, but differs from experimental results. For a correct treatment of the experiment, we indeed need to consider a quantum dot with two barriers where the situation is more complicated and analytical results can only be obtained in special cases. As  before, the second and higher cumulants at large times approach constant values which, however, numerically deviate from the values  (11), (12) for the single barrier case. We therefore performed  stochastic trajectory simulations (not shown) with the single dot parameters of the experiment. The simulations quantitatively reproduce the temporal freezing of the second cumulant $C_2$ in Fig.~4a, and the  temporal behavior $C^k_n(t)$ of the higher cumulants in Fig.~4e.

\end{document}